\documentclass[superscriptaddress,twocolumn,amsmath,amssymb,aps,prl]{revtex4-1}

\usepackage{amsmath}
\usepackage{graphicx}
\usepackage{dcolumn}
\usepackage{bm}
\usepackage{float}
\begin{document}

\title{Tunneling spectroscopy of localized states of WS$_2$ barriers in vertical van der Waals heterostructures}

\author{Nikos Papadopoulos}
\email{Email: n.papadopoulos@tudelft.nl\\}
\affiliation {\small \textit Kavli Institute of Nanoscience, Delft University of Technology, Lorentzweg 1, Delft 2628 CJ, The Netherlands}

\author{Pascal Gehring}
\email{Email: p.gehring@tudelft.nl\\}
\affiliation {\small \textit Kavli Institute of Nanoscience, Delft University of Technology, Lorentzweg 1, Delft 2628 CJ, The Netherlands}

\author{Kenji Watanabe}
\affiliation {\small \textit National Institute for Materials Science, 1-1 Namiki, Tsukuba 305-0044, Japan}

\author{Takashi Taniguchi}
\affiliation {\small \textit National Institute for Materials Science, 1-1 Namiki, Tsukuba 305-0044, Japan}

\author{Herre S. J. van der Zant}
\affiliation {\small \textit Kavli Institute of Nanoscience, Delft University of Technology, Lorentzweg 1, Delft 2628 CJ, The Netherlands}

\author{Gary A. Steele}
\affiliation {\small \textit Kavli Institute of Nanoscience, Delft University of Technology, Lorentzweg 1, Delft 2628 CJ, The Netherlands}




\
\begin{abstract}
In transition metal dichalcogenides, defects have been found to play an important role, affecting doping, spin-valley relaxation dynamics, and assisting in proximity effects of spin-orbit coupling. Here, we study localized states in WS$_2$ and how they affect tunneling through van der Waals heterostructures of h-BN/graphene/WS$_2$/metal. The obtained conductance maps as a function of bias and gate voltage reveal single-electron transistor behavior (Coulomb blockade) with a rich set of transport features including excited states and negative differential resistance  regimes. Applying a perpendicular magnetic field, we observe a shift in the energies of the quantum levels and information about the orbital magnetic moment of the localized states is extracted.  
\end{abstract}

\maketitle

\section{Introduction}
\medskip

\noindent
Tunneling spectroscopy in van der Waals and other heterostructures is a powerful tool that can reveal unique information about the density of states (DOS) of the electrodes \cite{Britnell2012,Jang2017}, about phonons (or other excitations) \cite{vdovin_phonon-assisted_2016,deVega2017,Ghazaryan2018}, about the chiral, valley \cite{wallbank_tuning_2016} and spin states of the carriers \cite{Eisenstein2017,Kim2018} and their interactions \cite{becker_probing_2011}. Recently it was shown that the presence of defects in crystalline hexagonal boron nitride (h-BN) tunneling barriers can be detected in the tunneling spectra, which is dominated by Coulomb blockade effects \cite{chandni_signatures_2016,Greenaway2018}. 

Semiconducting layered materials such as transition metal dichalcogenides (TMDCs) are promising building blocks for transistors and tunneling devices \cite{Radisavljevic2011,Ovchinnikov2014,Lin2015}. Furthermore, because of their crystallinity and absence of surface dangling bonds, they can be used as ideal substrates and barriers\cite{georgiou_vertical_2013,Kretinin2014,Dvir2018} when the Fermi level is placed inside their band gap. Earlier studies have shown that WS$_2$, which has the largest band gap among the TMDCs, is a promising material for tunneling transistors \cite{georgiou_vertical_2013,Li2017}. Moreover heterostructures of graphene and WS$_2$ are interesting for proximity induced spin-orbit coupling in graphene \cite{Avsar2014} as well as for tuning the excitonic properties of TMDCs \cite{Raja2017}.

Here, we study tunneling spectroscopy of h-BN/graphene/WS$_2$/metal heterostructures at low temperatures. The conductance maps show clear Coulomb diamonds (CDs), which originate from tunneling through localized states formed by defects in the WS$_2$ barriers. Such states have also been found in electrostatically defined WS$_2$ quantum dots\cite{Song2015} but have not been studied in detail. We find that these localized states have a radius on the order of 2-12~nm in the plane of the WS$_2$, in agreement with previous reports \cite{Song2015}. Finally, by studying the behavior of the energies of different charge transitions under a perpendicular magnetic field, we calculate their orbital magnetic moments. Using these values, we estimate the spatial extent of the individual states. 

\begin{figure}
  \begin{center}
    \includegraphics[width=8.7cm]{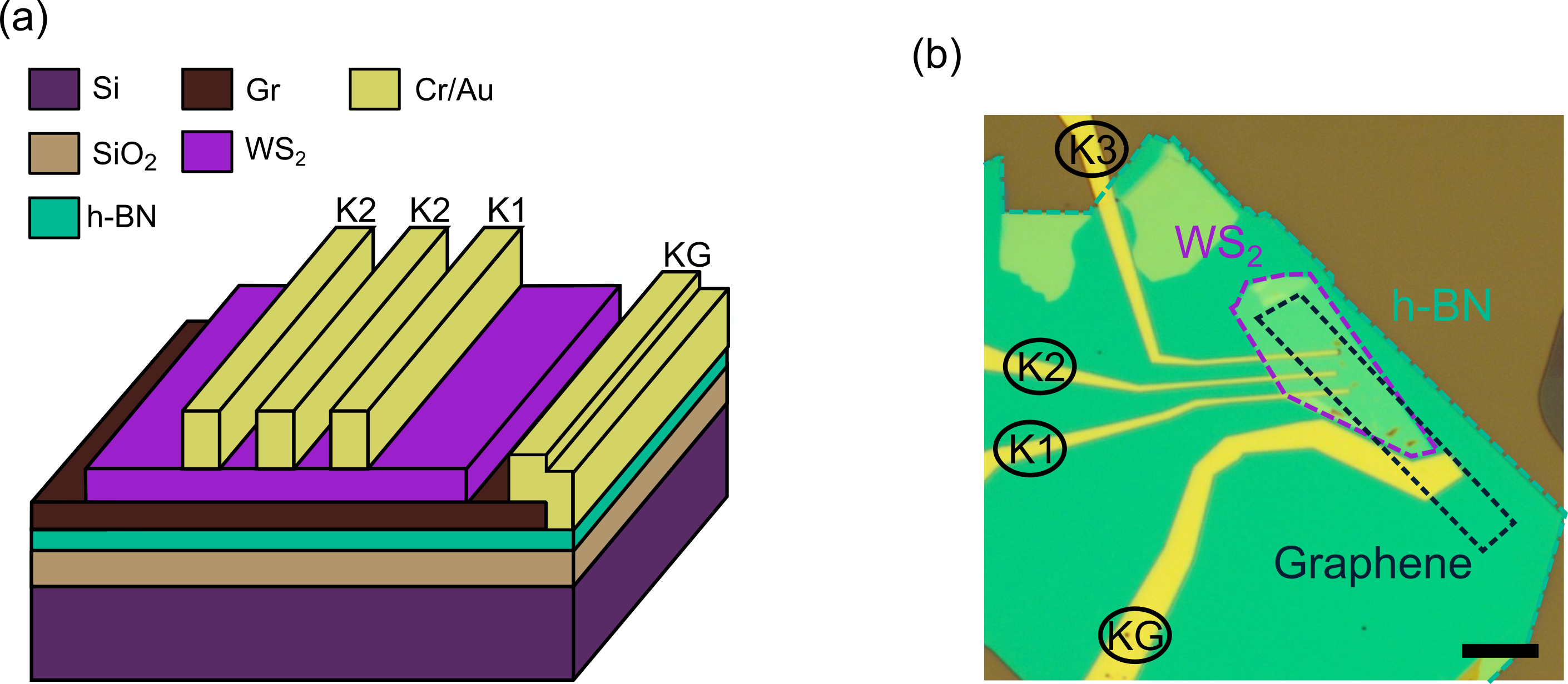}
      \caption{\small Device schematic and optical image. (a) 3D schematic illustration of a device illustrating the different layers and electrical contacts. (b) Optical image of a final device. The dotted purple and green lines indicate the shape of the WS$_2$ and h-BN flake, respectively. K1, K2 , K3 and KG are labels of the contacts. Scale bar: 5~$\mu$m.}
  \end{center}
\end{figure}

\section{Results and discussion}
\medskip

\begin{figure*}
  \begin{center}
    \includegraphics[width=15 cm]{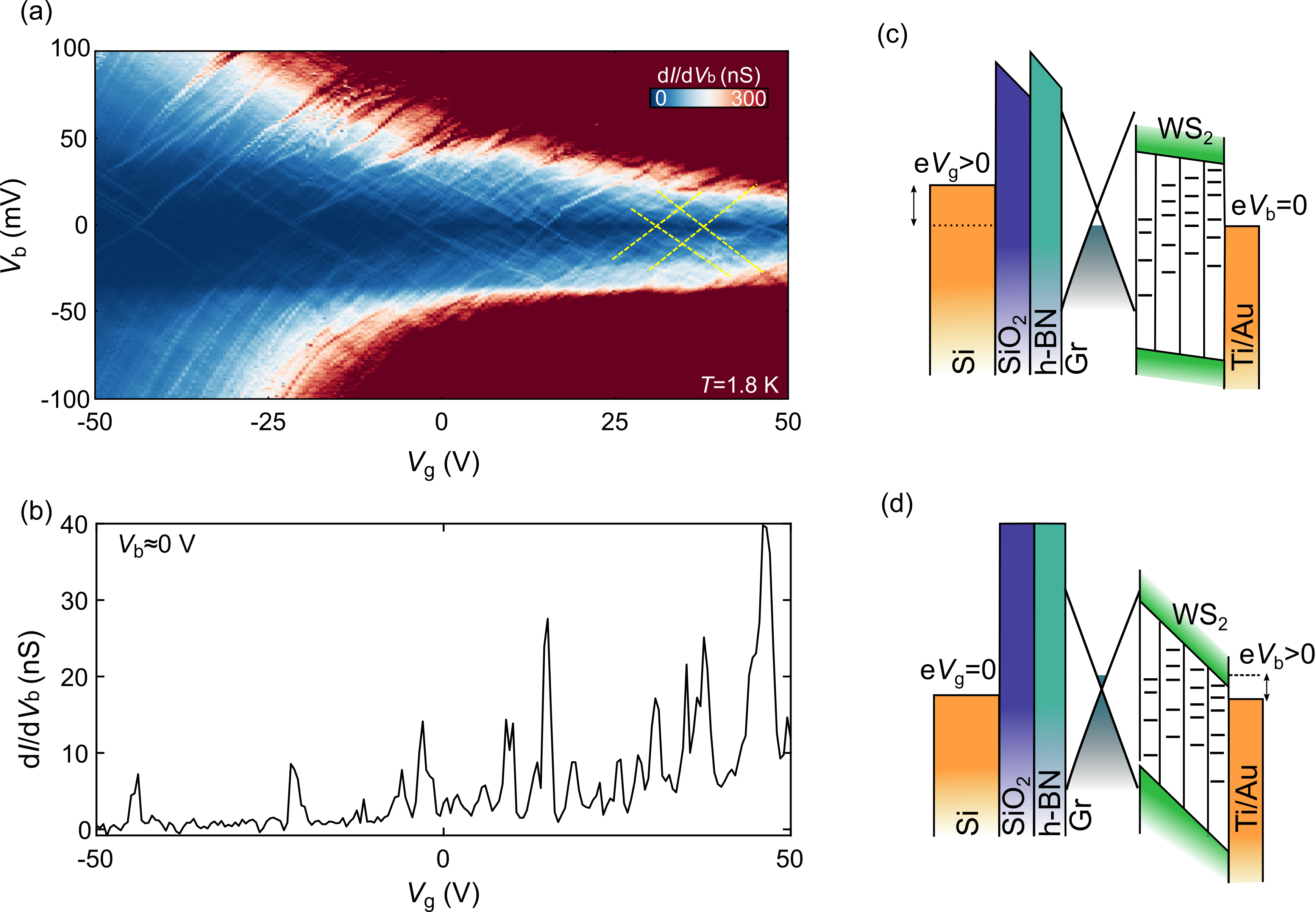}
      \caption{\small Tunneling spectroscopy at low temperatures. (a) Differential conductance ($\text{d}I/\text{d}V_{\text{b}}$) as a function of bias-voltage ($V_{\text{b}}$) and back-gate voltage ($V_{\text{g}}$) recorded at $T=1.8$~K. The green dashed lines indicate two Coulomb-diamonds. (b) Line-cut showing at $\text{d}I/\text{d}V_{\text{b}}$ versus $V_{\text{g}}$ at $V_{\text{b}}=0$~V extracted from (a). (c),(d) Energy band diagram of the vertical heterostructure under $V_{\text{g}}>$0~V and $V_{\text{b}}$=0~V in (c) and $V_{\text{g}}$=0~V and $V_{\text{b}}>$0~V in (d). The back-gate is capacitively coupled to the graphene as well as to the WS$_2$, which has discrete states in the band-gap that are located in different layers. Carrier tunneling between the two electrodes takes place via the impurity states. When the bias voltage $|V_{\text{b}}|$ increases above a certain point, carriers can directly tunnel into the conduction band of WS$_2$ thereby increasing the conductance significantly (regions in panel a).}
    \end{center}
\end{figure*} 


Devices with thin WS$_2$ tunnel barriers have been fabricated by a van der Waals pick-up method using poly-carbonate (PC) films \cite{zomer_fast_2014-1}. A four-layer WS$_2$ flake and a monolayer flake were chosen as tunneling barriers. The bottom electrode in the devices consists of a monolayer graphene, while for the top tunneling electrodes Cr/Au or graphite was deposited on the WS$_2$ flakes. Moreover, we employ hexagonal boron nitride (h-BN) as bottom dielectric with a thickness of 40~nm. The flat h-BN substrates eliminate extrinsic disorder in graphene and WS$_2$ \cite{Dean2010}. Figure 1a and 1b show a schematic illustration of the 4L-WS$_2$ device and an optical image of the heterostructure, respectively. All the measurements shown in the main text are data taken from the device with four layers of WS$_2$ (D1), by using the KG-K1 tunneling electrodes (see Fig. 1). Data from the other electrodes can be found in the supplemental material section. The device with monolayer WS$_2$ (D2) as barrier did not show any tunneling characteristics, an observation that is in-line with reports on monolayer MoS$_2$ Josephson junctions, where metallic behavior in the monolayers was observed \cite{island_thickness_2016}. Results from the D2 device can be found in the supplemental material.


Figure 2a shows the dependence of the differential conductance ($\text{d}I/\text{d}V_{\text{b}}$) on the tunnel bias ($V_{\text{b}}$) and back-gate voltage ($V_{\text{g}}$) at a temperature of 1.8~K. Although at room temperature the conductance of the device is a few $\mu$S (see Supplemental material), a significant decrease of three orders of magnitude in the values of $\text{d}I/\text{d}V_{\text{b}}$, as well as many resonant features from Coulomb blockade effect are evident. The observed Coulomb diamonds in the differential conductance originate from parallel tunneling through multiple states originating from impurities in the WS$_2$ layers (Fig. 2c), presumably sulfur or tungsten vacancies. Similar features in hexagonal boron nitride tunneling barriers have been reported recently \citep{chandni_signatures_2016, bretheau_tunnelling_2017, greenaway_tunnel_2018, Liu2018}. Moreover, an increase in the number of the resonances with back-gate voltage is observed (Fig. 2b), that can be attributed to the presence of localized states close to the edge of the conduction band, which could originate from sulfur vacancies similar to MoS$_2$ \cite{McDonnell2014}. Interestingly even at $V_\text{g}=50$~V, the zero-bias background conductance is suppressed, verifying that the Fermi level of WS$_2$ is below the conduction band edge. 

The conductance map in Fig. 2a provides further information about the nature of the localized states in the WS$_2$ sheet. The full-width-at-half-maximum (FWHM) of the zero-bias peaks is found to be close to $3.5k_\text{B}T$ which suggests that the system is in the thermally broadened, weak coupling regime i.e. the electronic coupling to the impurity states is the smallest energy scale. Furthermore, from the slopes of the diamonds we can extract their capacitive couplings to the leads ($C_{\text{s}}$, $C_{\text{d}}$) and the back gate ($C_{\text{g}}$) and can calculate the lever arm $\alpha=C_{\text{g}}/(C_{\text{s}}+C_{\text{d}}+C_{\text{g}})$ for the quantum dots \cite{Thijssen2008}. We find that $\alpha$ follows a distribution of values in the range of $0.8-2.5$~meV/V, with a maximum at 1.3$\pm$0.1~meV/V (see Fig. S2). Interestingly, we observe several diamonds with large values of $\alpha$ at large negative gate-voltages. A possible origin of these diamonds are different types of impurities with a smaller localization size and on-site capacitance. Assuming that the level spacing ${\Delta}E$ is much smaller than the charging energy $E_\text{c}$, such that the addition energy $E_\text{add} \approx 2E_\text{c}$, we can estimate the size of the quantum dots using $2E_\text{c}=\frac{e^2}{8\epsilon_0\epsilon_\text{r} r}$, for disk shape dots. Here $e$ is the electron charge, $\epsilon_0$, the vacuum permittivity, $r$ is the dot radius and $\epsilon_\text{r}=3.9$ the effective relative permittivity of SiO$_2$ and h-BN. Using the extracted values of $E_\text{add} = 50 - 75$~meV, this assumption yields a quantum dot size of about 9-12~nm. It is worth to mention that the diamonds and their positions change significantly after thermally cycling the device (see Supplemental Material).

\begin{figure}
  \begin{center}
    \includegraphics[width=8.7 cm]{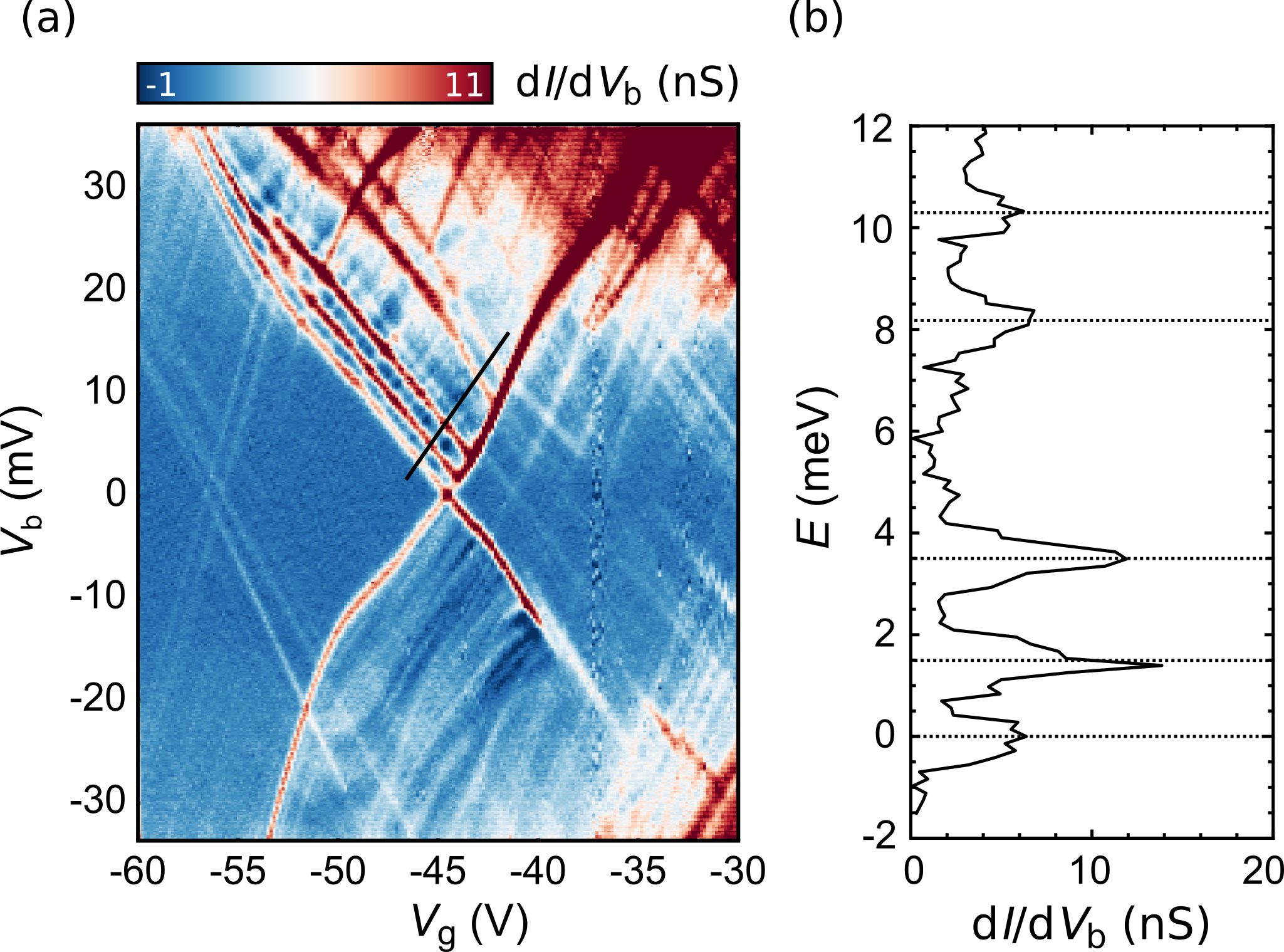}
      \caption{\small Excited state spectroscopy of defect states. (a) Detailed differential conductance map as a function of bias voltage $V_\text{b}$ and gate voltage $V_\text{g}$. The Coulomb diamond shows four excited state lines that run in parallel to the diamond edge. (b) Energy as a function of differential conductance from the linecut in panel (a). The high conductance peaks correspond to the excited states and have energies of 1.4, 3.5, 8.2 and 10.3~meV.  }
    \end{center}
\end{figure}

Above a voltage bias $|V_{\text{b}}|$ of 30~mV the conductance in the map of Fig. 2a increases significantly and many lines from diamonds can be observed. To explain such an increase of the current and the differential conductance at high bias, we have to take into account the effects of the interlayer voltage bias on the shape of the barriers. Although at zero bias and non zero gate voltage, the bias window contains only states within the band gap that assist the tunneling process (Fig. 2c), at higher bias the shape of tunneling barrier is modified and the WS$_2$ conduction band is lowered  \cite{Simmons1963, Sarkar2015}. This results in a field emission of the carriers above the band gap of WS$_2$ (Fig. 2d) \cite{georgiou_vertical_2013, Sarkar2015}, where carriers from the graphene and the electrode can tunnel directly into the WS$_2$ layer, which increases the conductance significantly.


Figure 3a shows a detailed differential conductance map of the gate voltage regime between -60 and -20~V, where a clear Coulomb diamond can be observed. This particular diamond exhibits excited states for positive bias and many negative differential resistance (NDR) features in the sequential tunneling regime. Excited states can have many different origins and can be phonon, spin or orbital excited states. Figure 3b shows  \textit{E} vs. $\text{d}I/\text{d}V_{\text{b}}$  obtained from the line-cut of Fig. 3a (black line). The energy axis has been shifted such that the the ground state is at 0~meV for clarity. From this we can extract the excited state energies of the first two excited states which are 1.4~meV and 3.5~meV, respectively. The next pair of excited states have energies of 8.2~meV and 10.3~meV. It is worth mentioning that the diamonds, apart from providing information about the localized states in WS$_2$, show features that originate from the graphene. In a few diamonds (like the one shown in Figure 4a), we observe lines with slopes that are different from those of the diamond edges. These lines represent multiple negative-differential-resistance (NDR) like feature. Earlier reports on molecular junctions using graphene electrodes \cite{gehring_distinguishing_2017} and tunneling spectroscopy through h-BN barriers \cite{zihlmann_non-equilibrium_2018}, have shown similar lines in their stability diagrams and they have been attributed to universal conductance fluctuations (UCFs) from the graphene contacts.  
 
\begin{figure*}
  \begin{center}
    \includegraphics[width=14 cm]{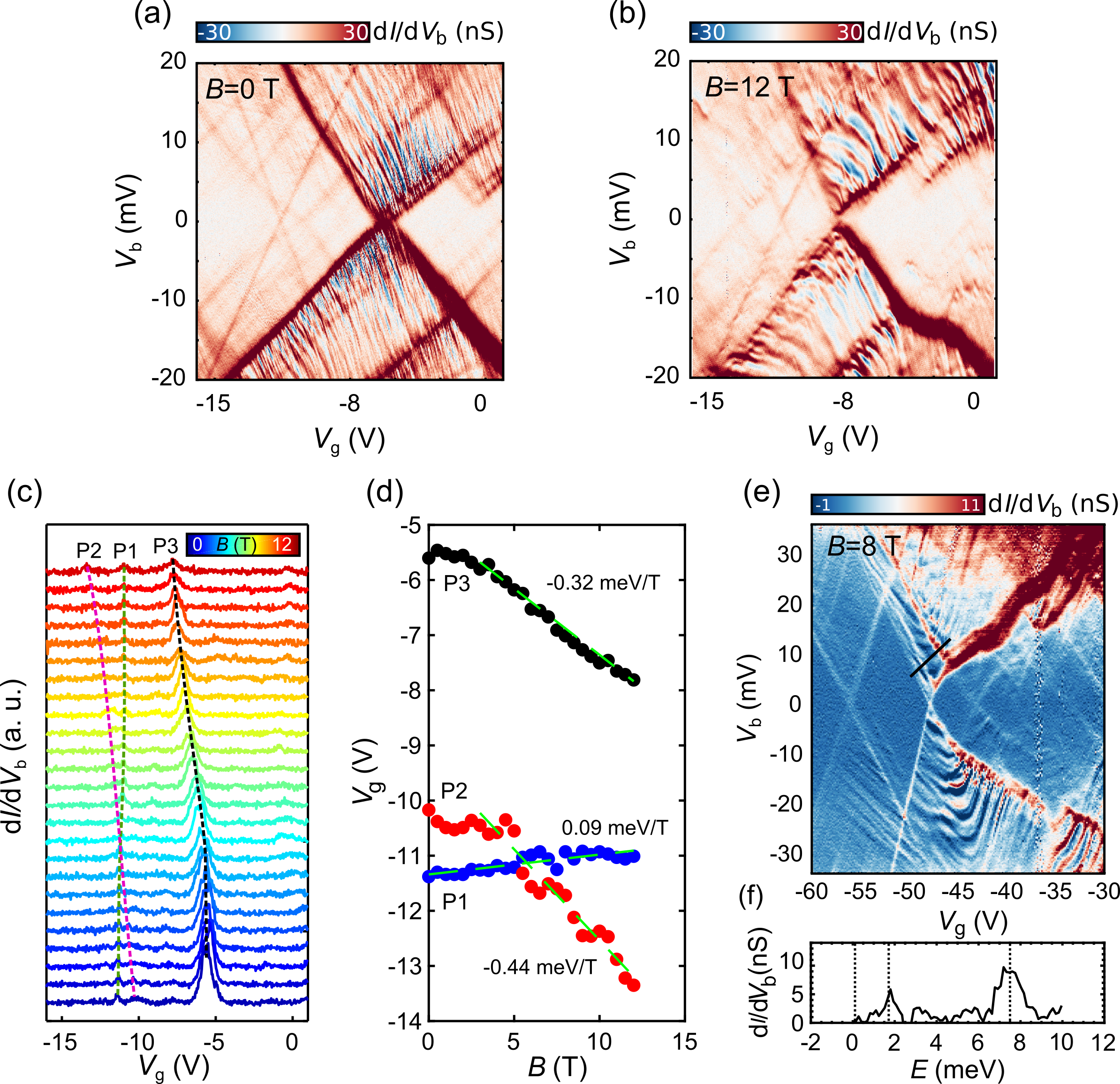}
      \caption{\small Ground state and excited state spectroscopy at finite magnetic fields. Differential conductance versus $V_{\text{b}}$ and $V_{\text{g}}$ obtained by applying $B$-fields of 0 in (a) and 12 T in (b). (c) Waterfall plot of $\text{d}I/\text{d}V_{\text{b}}$ \textit{vs.} $V_{\text{g}}$ for different magnetic fields. The dashed lines track the shifts of the three different resonances. (d) Position of the P1, P2 and P3 peaks in (c) as a function of magnetic field. From linear fits to the data we extract $\mu_B=\frac{\Delta E}{\Delta B}$ equal to 0.09$\pm$0.03, 0.44$\pm$0.05 and 0.32$\pm$0.03~meV/T for P1, P2 and P3 respectively. (e) $\text{d}I/\text{d}V_{\text{b}}$ map of Fig. 3 at $B$=8~T and $\text{d}I/\text{d}V_{\text{b}}$ $vs.$ $E$ (f) from the linecut of panel (e).}
    \end{center}
\end{figure*}


Upon application of a magnetic field perpendicular to the 2D plane, the Coulomb diamonds can reveal information about the spin and the orbital states. To this end we recorded the stability diagrams in the Fig. 3a and Fig. S1 at magnetic fields up to 12~T in order to study the field dependence of the ground and the excited states. Figure 4 shows the $\text{d}I/\text{d}V_{\text{b}}$ maps at 0 and 12~T, in (a) and (b), respectively. From the former, we observe clearly three zero-bias resonances, from which two have same values of $\alpha$ ($\sim$1.3~mV/V) while the other one has a slightly higher value of $\alpha \approx 2.1$~mV/V, presumably originating from a different type of defect. 
As it can be seen in Fig. 4b, the different types of diamonds have shifted horizontally at $B=12$~T, which can be explained by the coupling of the electron magnetic moment to the magnetic field.  Moreover, the UCF lines decrease in number upon increasing $B$ due to the formation of Landau levels in the DOS of the graphene. Additionally, we observe bending of the excited state lines as well as the edges of the Coulomb diamonds, a finding which could be explained by strong capacitive coupling between multiple dots that contribute to transport.\cite{Hofheinz2006}.

Figure 4c shows $\text{d}I/\text{d}V_{\text{b}}$ versus $V_{\text{g}}$ at zero bias, for magnetic fields from 0 to 12~T in steps of 0.5~T (blue to red). For $B=0$~T, the three ground-state resonance peaks are located at -7~V, -10~V and -11~V, which we label P1, P2 and P3, respectively. P1 and P3 have similar gate couplings, while P2 has a larger $\alpha$ as was seen in Fig. 4a. The P2 and P3 peaks shift in a similar way to more negative energies with increasing magnetic field, indicating that the two states have the same orbital wave-function. The P1 peak, on the other hand, shifts weakly to more positive gate values. Such shifts of the three states, originate from the coupling of the magnetic moments of the localized states to the external magnetic field. This becomes apparent when plotting the peak positions of P1, P2 and P3 as a function of magnetic field (Figure 4 (d)). The field dependence of the P2 and P3 peaks is non-linear, indicating a strong orbital magnetic moment of the involved ground state. Furthermore, from the linear fit to the high magnetic field data of peaks P1, P2 and P3, we get magnetic moments of 0.09$\pm$0.03, 0.44$\pm$0.05 and 0.32$\pm$0.03~meV/T, respectively. From the Zeeman energy, $E_z=\frac{1}{2} \mu B=\frac{1}{2} g \mu_{B} B$, where $g$ is the Land\'{e} \textit{g}-factor, $\mu$ and $\mu_{B}$ the magnetic moment and the Bohr magneton, \textit{g}-factors 3.4$\pm$0.9, 15.8$\pm$1.8, 11.0$\pm$0.9. All the corresponding $g$-factors are much larger than the value of 2 expected for a free electron. Thus, orbital magnetic moments play a significant role in the magnetic field dependent transport of defects in WS$_2$.

Finally, we recorded the stability diagram shown in Fig. 3 at a magnetic field of 8~T (Fig. 4e). Besides the aforementioned bending of the excited state lines and the edges of the Coulomb diamond we observe a drastic change of the excited state energies when comparing the data recorded at 0 and 8~T: out of four lines observed at 0~T (see Fig. 3), we only observe two lines with energies 1.8~meV and 7.3~mV at 8~T (Fig. 4f). Such strong changes of the excited state energies further suggests that the corresponding $g$ factors are large and/or that the spin of the excited state is $\gg 1/2$.

Unlike the hydrogen atom, whose ground-state has an orbital magnetic moment of zero ($m=0$) due to spherical symmetry, in the case of a dot formed in a 2D material there is a finite out-of-plane orbital magnetic moment even for the \textit{s}-shell, as a result of the reduced dimension along the $z$-axis. The orbital magnetic moment of a moving charge in a circular orbit, is given by \cite{minot_determination_2004,island_interaction-driven_2018}: $\mu_{\text{orb}}=rev_F/2$, where $r$ is the radius of the orbital, $v_{\text F}$ the Fermi velocity and $e$ the charge of the electron. Assuming a Fermi velocity around $1\times 10^5$~m/s, we obtain a radius for the P1, P2 and P3 resonances of 2$\pm$1, 9$\pm$2 and 6$\pm$1~nm, in relatively good agreement with the estimations of the dot size based on the capacitance extracted from the addition energies (Fig. 2a).

\section{Conclusions}
\medskip
In summary, we have performed tunneling spectroscopy measurements in van der Waals heterostructures with WS$_2$ barriers. Our data reveal a rich spectrum of Coulomb diamonds that we attribute to localized states inside the WS$_2$ sheet. These states localize carriers in a radius of 2-10~nm and show large orbital magnetic moments. Recent experiments indicate that such localized states have intrinsic spin-orbit coupling as large as 230~meV, and therefore tunneling devices as the ones shown here can be used for single spin-polarized electron injection. Lastly, localized states in TMDCs are important for single photon emission \cite{Aharonovich2016, Perebeinos2015} and if properly engineered could provide new systems for quantum communications, similar to nitrogen-vacancy centers in diamond.

\begin{center}
  \textbf{\small ACKNOWLEDGMENTS}
\end{center}

\noindent
This work is part of the Organization for Scientific Research (NWO) and the Ministry of Education, Culture, and Science (OCW). P.G. acknowledges a Marie Skłodowska-Curie Individual Fellowship under Grant TherSpinMol (ID: 748642) from the European Union’s Horizon 2020 research and innovation programme. Growth of hexagonal boron nitride crystals was supported by the Elemental Strategy Initiative conducted by the MEXT, Japan and the CREST (JPMJCR15F3), JST.


\bibliography{WS2_tunneling_references}

\end{document}